\documentclass[lettersize, journal]{IEEEtran}
\usepackage{lineno,hyperref}
\usepackage{amsmath}
\usepackage[ruled, lined, longend, linesnumbered]{algorithm2e}
\usepackage[square,numbers]{natbib}
\usepackage{amsmath, amsthm, amssymb}

\usepackage{amsthm}
\usepackage{placeins}
\usepackage{csquotes}
\usepackage{url}
\usepackage{multirow}
\usepackage{amssymb}
\usepackage{graphics}
\usepackage[utf8]{inputenc}
\usepackage[T1]{fontenc}
\usepackage{subcaption}
\usepackage{caption}
\usepackage{algpseudocode}
\usepackage{amsfonts}
\usepackage{filecontents}
\usepackage{mathtools}
\usepackage{booktabs}
\usepackage{enumitem}
\usepackage{epstopdf}
\setlist[itemize]{leftmargin=*}

\usepackage{xcolor}
\definecolor{rv1}{rgb}{1.0, 0.44, 0.37}
\definecolor{rv2}{rgb}{0.4, 1.0, 0.0}
\definecolor{rv3}{rgb}{0.0, 0.75, 1.0}
\definecolor{rvt}{rgb}{0.75, 0.75, 0.75}
\usepackage{soul}

\newtheoremstyle{exampstyle}
{7pt} 
{7pt} 
{\itshape} 
{} 
{\bfseries} 
{.} 
{.5em} 
{} 
\theoremstyle{exampstyle}

\algrenewcommand\algorithmicindent{0.5em}%
\ifCLASSINFOpdf
\graphicspath{{Figures/}}
\else
\fi
\newsavebox\mybox

\begin{document}
	\title{GNN-Based Polarforming for Multi-User MISO Short-Packet URLLC under Imperfect CSI}
	\author{\IEEEauthorblockN{~Zahra~Mehrzad,~Hamed~Aghaei-Karkaj,~Rahman~Saadat~Yeganeh,~Kamran~Ebrahimi, Mohammad~Robat~Mili,~Symeon~Chatzinotas,~\textit{Fellow, IEEE},~and~Ioannis~Krikidis,~\textit{Fellow, IEEE}
		}
		\thanks{Z. Mehrzad, and M. Robat Mili are with the Pasargad Institute for Advanced Innovative Solutions (PIAIS), Tehran, Iran (email: \{zahra.mehrzad, mohammad.robatmili\}@piais.ir). H. Aghaei-Karkaj and R. Saadatyeganeh are with the Department of Electrical Engineering, Sharif University of Technology, Tehran, Iran (email: \{Hamed.aghaei78, rahman.saadat\}@sharif.edu). K. Ebrahimi is with the Department of Electrical Engineering, Amirkabir University of Technology, Tehran, Iran (email: kamranebrahimi@aut.ac.ir). S. Chatzinotas is with the Interdisciplinary Centre for Security, Reliability and Trust (SnT), University of Luxembourg, L-1855 Luxembourg City, Luxembourg (email: symeon.chatzinotas@uni.lu). I. Krikidis is with the Department of Electrical and Computer Engineering, University of Cyprus, Nicosia 1678, Cyprus (email: krikidis.ioannis@ucy.ac.cy).}
	}
	\maketitle
\begin{abstract}
This paper investigates polarization-aware transmission for multi-user multiple-input single-output (MU-MISO) short-packet ultra-reliable low-latency communications (URLLC) under imperfect channel state information (CSI). We consider a system in which the base station (BS) and users are equipped with polarization-reconfigurable antennas that enable adaptive polarization states through controllable polarization coefficients. A multi-objective optimization problem is formulated to jointly maximize the finite-blocklength (FBL) achievable sum rate and minimize the maximum decoding error probability (DEP), subject to transmit-power, latency, reliability, and discrete polarization-control constraints. The resulting multi-objective problem is scalarized using a normalized weighted-sum utility. To enable low-complexity online decision-making, a heterogeneous graph neural network (GNN) is developed to learn the joint mapping from estimated polarized CSI to digital beamforming and transmit/receive polarforming vectors (PFVs) while accounting for the system constraints. Numerical results demonstrate that the proposed GNN-based polarforming (PF) framework substantially improves the FBL sum rate while reducing the maximum DEP compared with conventional fixed-polarization schemes, particularly under channel depolarization and imperfect CSI. This highlights the potential of adaptive polarization control for reliable low-latency transmission.
\end{abstract}

\vspace{-1em}
\section{Introduction}
\IEEEPARstart{U}{RLLC} is a key service category for beyond-5G and 6G networks, supporting mission-critical applications that require extremely low packet error probability and latency~\cite{3GPP2016,Vu2025SPC}. In such applications, short-packet transmission is essential. In the finite-blocklength (FBL) regime, classical Shannon capacity is insufficient because the achievable rate depends jointly on the SINR, blocklength, and channel dispersion, resulting in an inherent rate reliability latency tradeoff~\cite{Polyanskiy2010,Ostman2021URLLC,Wu2026FBL}. This motivates transmission designs that explicitly account for decoding error probability (DEP) rather than optimizing rate alone.

Multi-user multiple-input single-output (MU-MISO) transmission can exploit spatial degrees of freedom through digital beamforming, but inter-user interference and the nonlinear FBL rate expression make optimization challenging under stringent URLLC constraints. Existing FBL-URLLC studies have shown that channel estimation accuracy and spatial processing can substantially affect reliability under imperfect CSI~\cite{Ostman2021URLLC,Peng2025URLLC}. Related multi-user wireless systems have also considered joint optimization under stringent user-level constraints~\cite{Yeganeh2025ASRIS}.

Polarization provides an additional electromagnetic degree of freedom that can complement conventional spatial processing. Depolarization and polarization mismatch caused by wireless propagation can degrade the received signal quality and impair transmission reliability, with the degree of depolarization characterized by the cross-polarization discrimination (XPD)~\cite{Zhou2024Polarforming,Ding2026Opportunities}. PF enables the transmitter and receiver to adjust their polarization configurations through phase-shifter-based polarization-reconfigurable antennas, supporting linear, circular, and elliptical polarization states~\cite{Zhou2024Polarforming,Zhou2025PS,Shao2026PAISAC,11547205}. Such adaptive polarization control can be particularly beneficial for URLLC systems, where transmission reliability is a key design requirement.

Recent studies have extended PF to integrated sensing and communication, multi-user and intelligent-surface-assisted communications, and LEO satellite systems~\cite{Shao2026PAISAC}-\cite{Li2026LEOPolarforming}. However, the role of polarization adaptability in short-packet URLLC, where reliability is governed by the FBL regime and is highly sensitive to channel quality, remains largely unexplored. In particular, the joint consideration of FBL-URLLC under imperfect CSI and the coupled design of digital beamforming with transmit and receive PF has not been systematically investigated. Moreover, iterative PF optimization may require repeated updates for different channel realizations, motivating low-complexity learning-based solutions~\cite{Zhou2024Polarforming,Zhou2025PS,Zhou2025ThreeD}.

Learning-based optimization has recently been explored for high-dimensional non-convex wireless problems under imperfect CSI~\cite{Yeganeh2025LEODRL}. GNNs are particularly attractive for multi-user wireless systems because they can represent user interactions and interference relationships through graph-based message passing. GNN-based beamforming has been investigated for MU-MISO systems~\cite{Li2024GNNMUMISO}, while heterogeneous GNNs have been applied to joint active/passive beamforming in distributed STAR-RIS-assisted systems~\cite{Le2025GNNSTAR}. Recent work has further shown that customized GNN objectives can jointly address communication performance and user-specific constraints~\cite{Tang2025TwoPhaseGNN}. Nevertheless, their use for polarization-aware FBL-URLLC transmission remains unexplored.

Motivated by these gaps, we investigate a MU-MISO short-packet URLLC system with polarforming antennas (PAs) at the BS and users under imperfect CSI. The proposed framework jointly optimizes digital beamforming, polarforming vectors (PFVs), and blocklength allocation to balance the FBL sum rate and maximum DEP subject to practical system constraints.
The main contributions are summarized as follows:

\begin{itemize}

\item We develop a polarization-aware MU-MISO framework for short-packet URLLC under imperfect CSI, incorporating an XPD-based polarized channel model and finite-resolution polarization control.

\item We formulate a multi-objective optimization problem that jointly designs digital beamforming and transmit/receive PFVs, and scalarize it through a normalized weighted-sum utility to provide a controllable tradeoff between FBL sum rate and maximum DEP.

\item We propose a heterogeneous GNN architecture with tailored message-passing and readout layers that learns the mapping from estimated polarized CSI to beamforming and PF variables, with constraint-aware output mappings and a customized training objective.

\item Numerical results demonstrate that adaptive PF improves the FBL sum rate and reduces the maximum DEP relative to fixed-polarization schemes, particularly under depolarization and imperfect CSI.

\end{itemize}
\vspace{-0.7em}
\section{System Model and Problem Formulation}
\begin{subsection}{System Model}	
We consider the downlink of a MU-MISO short-packet communication system. The BS is equipped with $M$ PAs serving $K$ single-antenna URLLC users, indexed by $\mathcal{K}=\{1,\ldots,K\}$, with the set of BS antennas denoted as $\mathcal{M}=\{1,\ldots,M\}$.
Both the BS and users employ PAs, each comprising a single RF chain connected to two orthogonally polarized antenna elements corresponding to vertical ($\mathcal{V}$-element) and horizontal ($\mathcal{H}$-element) polarizations, respectively, through tunable attenuators and a phase shifter. By independently adjusting the amplitudes of the two components and their relative phase, each PA can synthesize arbitrary polarization states. This capability enables adaptive polarization-domain beam control, allowing the transceiver pair to adjust its polarization states according to the effective polarized channel, thereby mitigating polarization mismatch, exploiting polarization diversity, and improving link reliability~\cite{Zhou2025PS}.

Let $\eta_{1}, \eta_{2} \in[0,1]$ denote the amplitude control factors for the $\mathcal{V}$- and $\mathcal{H}$-polarized components of a PA, respectively, and let $\phi \in[0,2 \pi)$ represent their relative phase shift. To describe the polarization characteristics of each PA, we define the PFV based on the combined amplitude and phase adjustments. Specifically, the transmit PFV for the $m$-th antenna element at the BS is given by

\vspace{-0.7em}
\begin {equation}
\mathbf{c}_m=\frac{1}{\sqrt{2}}\left[\eta_{1,m}, \eta_{2,m}e^{-j \phi_{m}}\right]^T,\quad \forall m \in \mathcal{M}.
\vspace{-0.5em}
\end {equation}

Correspondingly, the receive PFV of user $k$ is defined as
\vspace{-0.7em}
\begin {equation}
\mathbf{w}_k=\left[\eta_{1,k}, \eta_{2,k}e^{-j \phi_{k}}\right]^T,\quad \forall k \in \mathcal{K}.
\vspace{-0.5em}
\end {equation}

In practice, the amplitude and phase control of each PFV can be implemented using finite-resolution digital quantization. For ease of implementation, we consider discrete amplitude and phase adjustments at both the BS and the users. Let $B_\eta$ and $B_\phi$ denote the number of quantization bits for amplitude and phase control, respectively.
The phase shift is selected from a finite quantization set over $[0,2\pi)$ with $2^{B_\phi}$ levels, i.e.,
$
	\phi \in \mathcal S \triangleq \left\{ 0, \frac{2\pi}{2^{B_\phi}}, \ldots, \frac{2\pi(2^{B_\phi}-1)}{2^{B_\phi}} \right\}.
$
Similarly, the amplitude coefficients are selected from a finite set of $2^{B_\eta}$ uniformly spaced levels in $[0,1]$, i.e.,
$
	\eta_{i} \in \mathcal A \triangleq \left\{ \rho_1, \rho_2, \ldots, \rho_{2^{B_\eta}} \right\}, i \in \{1,2\},
$
where $0 \leq \rho_1 < \cdots < \rho_{2^{B_\eta}} \leq 1$. 

We consider a narrowband quasi-static block fading channel model, where the effective polarformed channel from the BS to user $k$ is determined by the interaction between the transmit and receive PFVs via their amplitude and phase configurations. The overall effective channel vector is given by
\vspace{-0.7em}
\begin{equation}
\mathbf{h}_{k}=\left[\mathbf{w}_k^H \mathbf{A}_{k,1} \mathbf{c}_1, \ldots, \mathbf{w}_k^H \mathbf{A}_{k,M} \mathbf{c}_M\right]^T \in \mathbb{C}^{M \times 1},
\vspace{-0.7em}
\end{equation} 
where $\mathbf{A}_{k,m} \in \mathbb{C}^{2 \times 2}$ denotes the polarized channel matrix between the BS $m$-th PA and user $k$. It is modeled as \cite{Zhou2025PS}
\vspace{-0.5em}
\begin{equation}
\mathbf{A}_{k,m}=\frac{1}{\sqrt{\chi_k+1}}\left[\begin{array}{cc}
	1 & \sqrt{\chi_{k}}  \\
	\sqrt{\chi_{k}} & 1
\end{array}\right] \odot \tilde{\mathbf{H}}_{k, m},
\vspace{-0.5em}
\end{equation} 
where $\chi_{\mathrm{k}}$ denotes the inverse XPD. The entries of $\tilde{\mathbf{H}}_{k,m}$ are i.i.d. circularly symmetric complex Gaussian (CSCG) random variables as $[\tilde{\mathbf{H}}_{k,m}]_{i,j}\sim \mathcal{CN}\left(0,\frac{\rho_k^2}{2}\right)$, where $\rho_k^2 = C_0 d_k^{-\alpha}$ captures the large-scale path loss, in which $d_k$ denotes the distance between the BS and user $k$, $\alpha$ is the path-loss exponent, and $C_0$ represents the reference path loss at a distance of 1 meter.

CSI plays a crucial role in PF systems, as its accuracy directly affects the ability to adapt polarization states and compensate channel depolarization. However, acquiring accurate CSI in practice is challenging. The hardware constraints inherent to PF architectures render conventional channel estimation techniques less effective \cite{Zhou2024Polarforming}. In addition, the strict latency requirements of URLLC systems make frequent and high-overhead channel training impractical, thereby rendering perfect CSI (PCSI) acquisition infeasible.
Therefore, the estimated BS-user channel is modeled as
$\hat{\mathbf{h}}_k = \mathbf{h}_k + \Delta \mathbf{h}_k,$
where $\Delta \mathbf{h}_k$ represents the channel estimation error and is assumed to follow a CSCG distribution, i.e., $\Delta \mathbf{h}_k \sim \mathcal{CN}(\mathbf{0}, \xi \mathbf{I}_M),$ where $\xi$ denotes the normalized variance of the channel estimation error.

The received signal at user $k$ is expressed as
$y_k = \hat{\mathbf{h}}_k^H \sum_{j=1}^{K} \mathbf{p}_j x_j + n_k,$
where $x_k \sim \mathcal{CN}(0,1)$ denotes the transmitted information symbol with unit power, and $\mathbf{p}_k \in \mathbb{C}^{M \times 1}$ represents the dedicated beamforming vector for user $k$. Moreover, $n_k \sim \mathcal{CN}(0,\sigma^2)$ is the additive white Gaussian noise (AWGN) at user $k$ with variance $\sigma^2$. Given the BS transmit power budget $P_{\text{BS}}^{\max }$, it follows that the total transmit power must satisfy $\sum_{k \in \mathcal{K}} \|\mathbf{p}_k\|^2 \leq P_{\text{BS}}^{\max}$.

Hence, the signal-to-interference plus-noise ratio (SINR) at user $k$ is calculated as \vspace{-0.4em}
\begin{equation}
	\gamma_{k}=\frac{\left|\hat{\mathbf{h}}_k^H\left(\mathbf{c}_m, \mathbf{w}_k\right) \mathbf{p}_k\right|^2}{\sum_{j \neq k}\left|\hat{\mathbf{h}}_k^H\left(\mathbf{c}_m, \mathbf{w}_k\right) \mathbf{p}_j\right|^2+\sigma^2}, \forall k \in \mathcal{K}.
\vspace{-0.4em}
\end{equation}

The FBL coding framework is adopted to characterize the achievable rate under a nonzero DEP. Specifically, the achievable rate of user $k$ is given by
\vspace{-0.4em}
\begin{equation}
	R_k=
	\log_2(1+\gamma_k)
	-
	\sqrt{\frac{V(\gamma_k)}{L_k}}
	Q^{-1}(\epsilon_k)\log_2 e ,
\vspace{-0.4em}
\end{equation}
where $\epsilon_k$ denotes the DEP, $L_k=WT_k$ is the blocklength in channel uses with $W$ denoting the system bandwidth and $T_k$ the transmission latency, and $Q^{-1}(\cdot)$ represents the inverse Gaussian $Q$-function. The channel dispersion is given by
$
	V(\gamma_k)=1-\frac{1}{(1+\gamma_k)^2}.
$
For a packet containing $D_k$ information bits, the corresponding DEP can be expressed as
\vspace{-0.3em}
\begin{equation}
	\epsilon_k=
	Q\left(
	\ln(2)
	\sqrt{\frac{L_k}{V(\gamma_k)}}
	\left(
	\log_2(1+\gamma_k)-\frac{D_k}{L_k}
	\right)
	\right).
	\vspace{-0.3em}
\end{equation}
\vspace{-0.5cm}
\end{subsection}
\vspace{-0.1cm}
\subsection{Problem Formulation}
The FBL rate and DEP are inherently coupled, yielding a fundamental rate--reliability tradeoff. Accordingly, we formulate the following multi-objective optimization problem under imperfect CSI to jointly maximize the achievable sum rate while minimizing the maximum DEP among all users.
\vspace{-0.5em}
\begin{subequations}
	\label{P1}
	\begin{align}
		\mathcal{P}_1:\quad
		&\max_{\{\mathbf{p}_k\},\{\mathbf{c}_m\},\{\mathbf{w}_k\}, \{L_k\}}
		\quad 
		\textstyle\sum_{k\in\mathcal K}R_k,
		\label{P1a}
		\\
		&\min_{\{\mathbf{p}_k\},\{\mathbf{c}_m\},\{\mathbf{w}_k\}, \{L_k\}}
		\quad 
		\max_{k\in\mathcal K}\epsilon_k,
		\label{P1b}
		\\
		&~~\text{s.t.}\quad
		\textstyle\sum_{k\in\mathcal K}\|\mathbf p_k\|^2
		\le
		P_{\mathrm{BS}}^{\max},
		\label{P1c}
		\\
		&~~~~~~~~~
		\epsilon_k\le\epsilon_{\max},
		\quad
		\forall k\in\mathcal K,
		\label{P1d}
		\\
		&~~~~~~~~~
		T_k\le T_{\max},
		\quad
		\forall k\in\mathcal K,
		\label{P1e}
		\\
		&~~~~~~~~~
		\phi\in\mathcal S,\;
		\eta_{i}\in\mathcal A,\;
		i\in\{1,2\},
		\label{P1f}
		\\
		&~~~~~~~~~
	    L_k\in\mathbb{Z}_{+}.
		\label{P1g}
			\vspace{-0.5em}
	\end{align}
\end{subequations}
Constraint \eqref{P1c} limits the BS transmit power, while \eqref{P1d} and \eqref{P1e} impose the URLLC reliability and latency requirements, respectively. Constraint \eqref{P1f} accounts for the finite-resolution PFV control, and \eqref{P1g} ensures an integer number of channel uses.
To obtain a tractable problem while providing a flexible tradeoff between throughput and reliability, we adopt the normalized weighted-sum approach. Specifically, 
	\vspace{-0.4em}
\begin{equation}
	U_1 = \frac{\sum_{k\in\mathcal K}R_k}{R_{\text{ref}}}, \qquad U_2 = \frac{\epsilon_{\text{ref}}-\max_{k\in\mathcal K}\epsilon_k}{\epsilon_{\text{ref}}},
	\vspace{-0.5em}	
	\end{equation}
where $R_{\text{ref}}$ and $\epsilon_{\text{ref}}$ are the corresponding normalization factors. Accordingly, the resulting single-objective normalized utility function is defined as
	\vspace{-0.6em}
\begin{subequations}
	\label{P2}
	\begin{align}
		\mathcal{P}_2:\quad
		\max_{\{\mathbf{p}_k\},\{\mathbf{c}_m\},\{\mathbf{w}_k\}, \{L_k\}}
		\mathcal{U}=\omega U_1+(1-\omega)U_2
		\label{P2a}\\
		\text{s.t.}\quad
		\eqref{P1c}-\eqref{P1g},
		\label{P2b}
			\vspace{-0.9em}
	\end{align}
\end{subequations}
where $0\leq\omega\leq1$ is the weighting factor. Problem $\mathcal{P}_2$ is non-convex due to the coupled dependence of
the FBL rate and DEP on the beamforming, PFVs, and blocklength, together with the discrete PFV constraints. Moreover, solving it iteratively for different channel realizations can incur substantial computational overhead. To enable low-complexity online optimization, we therefore develop the proposed GNN-based solution in Section~III.
	\vspace{-0.5em}
\section{Proposed GNN-Based Solution Strategy}
In this section, we present the proposed graph neural network (GNN) that
learns the mapping from the CSI of each user
directly to the digital beamforming vectors, the transmit/receive
PFVs, and the blocklength allocation, in order to
solve problem $\mathcal{P}_2$. We
first describe the graph representation of the system, then detail the
three building blocks of the network: the \emph{initial layer}, the
\emph{node-update layer}, and the \emph{readout layer}. Finally
present the training objective.
	\vspace{-1.2em}
\subsection{Graph Representation}
The proposed GNN represents the system as a heterogeneous graph with
$K+1$ nodes: $K$ user nodes, indexed by $k \in \mathcal{K}$, and a single
BS node. Every user node $k$ is connected to the BS node with an edge whose
weight reflects the quality of user $k$'s link, $e_k = \lVert \hat{\mathbf{h}}_k \rVert_2, \forall k \in \mathcal{K}$ (i.e., the norm of the estimated effective polarized channel of user $k$.). This scalar is used to weight the contribution of
each user when aggregating information at the BS node. No explicit user-to-user edge features are learned; instead, the interaction among users is captured through a symmetric mean aggregation over all user embeddings, so that every user node is implicitly and uniformly connected to every other user node.
\vspace{-1.8em}
\subsection{Architecture of the Proposed GNN}
At each user node $k$, the associated optimization variables, including the beamforming vectors, the transmit and receive PFVs and blocklength, are updated simultaneously across all nodes. Meanwhile, the PFM is derived at the BS node. Furthermore, we assume that every user node can share its local information to the BS. The node operations and information exchange are presented in three layers, as detailed below.
\subsubsection{Initial Layer}
The initial layer maps the raw per-user feature $\mathbf{x}_k = [\Re\{\hat{\mathbf{h}}_k\},\, \Im\{\hat{\mathbf{h}}_k\}] \in \mathbb{R}^{2M}$ to the
initial user embedding through a single feed-forward branch $\mathbf{v}_k^{(0)} = f_u(\mathbf{x}_k)$, where $f_u(\mathbf{x}_k)\in \mathbb{R}^{q}$ is feature extractor. This feature enables the data to be prepared for subsequent processing stages and message passing among nodes. The edge weights are $\ell_2$-normalized across the
user dimension as $\tilde{e}_k = \frac{e_k}{\sqrt{\sum_{j \in \mathcal{K}} e_j^2}}, \forall k \in \mathcal{K}$,
and used to compute a channel-quality-weighted input as $\mathbf{x}_{bs}^{(0)} = \frac{\sum_{k \in \mathcal{K}} \tilde{e}_k\, \mathbf{v}_k^{(0)}}{\sum_{k \in \mathcal{K}} \tilde{e}_k}$. The initial BS embedding is then obtained through a second feed-forward branch $\mathbf{v}_{bs}^{(0)} = f_{bs}(\mathbf{x}_{bs}^{(0)}) \in \mathbb{R}^{q}$, where $f_{bs}(\mathbf{x}_{bs}^{(0)})$ is the coresponding initial BS feature extractor.

\subsubsection{Node Messaging and Update Layer}
Once the initial features have been extracted, a multi-layer architecture is adopted to allow for message exchange between nodes and facilitate message passing among nodes and enable information sharing across the network. For $\ell \in \{1, \dots, L\}$, the node embeddings are refined by aggregating information from the previous layer. First, an unweighted mean of the user embeddings summarizes the state of all users as $\bar{\mathbf{v}}^{(\ell-1)} = \frac{1}{K}\sum_{j \in \mathcal{K}} \mathbf{v}_j^{(\ell-1)}$, and the channel-quality-weighted aggregation is recomputed using the layer-$(\ell-1)$ user embeddings:
$
	\mathbf{x}_{bs}^{(\ell-1)} = \frac{\sum_{k \in \mathcal{K}} \tilde{e}_k\, \mathbf{v}_k^{(\ell-1)}}{\sum_{k \in \mathcal{K}} \tilde{e}_k}.
$

The BS embedding is updated by $f_{bs}^{(\ell)}: \mathbb{R}^{3q} \to \mathbb{R}^{q}$, applied with a scaled residual connection as 
	$\mathbf{v}_{bs}^{(\ell)} = f_{bs}^{(\ell)}([\mathbf{v}_{bs}^{(\ell-1)},\, \bar{\mathbf{v}}^{(\ell-1)},\, \mathbf{x}_{bs}^{(\ell-1)}]) + \tau \, \mathbf{v}_{bs}^{(\ell-1)}$.
Similarly, each user embedding is updated by $f_u^{(\ell)}: \mathbb{R}^{2q} \to \mathbb{R}^{q}$, also with a scaled residual
connection. The user embedding is updated by $\mathbf{v}_k^{(\ell)} = f_u^{(\ell)}([\mathbf{v}_k^{(\ell-1)},\, \bar{\mathbf{v}}^{(\ell-1)}]) + \tau\, \mathbf{v}_k^{(\ell-1)}, \forall k \in \mathcal{K}.$

The equations are applied one after another over $L$ layers, with each layer having its own trainable parameters $\{f_u^{(\ell)}, f_{bs}^{(\ell)}\}_{\ell=1}^{L}$. This layered structure ensures that information is gradually shared between the BS node and every user node, while the embedding size remains constant at $q$.
\subsubsection{Readout Layer}
After $L$ message-passing layers, the final user and BS embeddings
$\mathbf{v}_k^{(L)}, \mathbf{v}_{bs}^{(L)} \in \mathbb{R}^{q}$ are mapped
to the optimization variables of problem $\mathcal{P}_2$ as follows.
\begin{itemize}
	\item \textbf{The beamforming vector}: For each user $k$, a linear readout
	head produces
	$\mathbf{p}_k^{(nn)} = \Psi_p(\mathbf{v}_k^{(L)}) \in \mathbb{R}^{2M}$,
	from which the complex beamforming vector is formed as
	$\bar{\mathbf{p}}_k = [p_{k,1}^{(nn)}, \dots, p_{k,M}^{(nn)}]^T + \jmath \cdot [p_{k,M+1}^{(nn)}, \dots, p_{k,2M}^{(nn)}]^T$.
	To satisfy the power constraint \eqref{P1b}, the beamforming vectors are
	jointly rescaled as $\mathbf{p}_k = \bar{\mathbf{p}}_k \sqrt{\frac{P_{\mathrm{BS}}^{\max}}{\sum_{j \in \mathcal{K}} \|\bar{\mathbf{p}}_j\|^2}},  \forall k \in \mathcal{K}.$
	\item \textbf{The blocklength ratio}:
	A second linear head followed by a sigmoid
	activation produces a normalized blocklength ratio per user,
	$L_k^{(nn)} = \mathrm{Sigmoid}\big(\Psi_L(\mathbf{v}_k^{(L)})\big) \in (0,1)$.
	As specified in constraint~\eqref{P1e}, the maximum packet length can be expressed as $L_{\max} = \lfloor T_{\max} W\rfloor$. Consequently, by discarding the random component of the output and setting the predicted length as $L_k^{(nn)} \cdot L_{\max} + 1$, both constraints~\eqref{P1e} and~\eqref{P1g} can be readily satisfied.
	\item \textbf{The transmit PFV}: For the transmit PFV, two linear heads are applied to the BS embedding
	$\mathbf{v}_{bs}^{(D)}$: $\mathbf{c}_1 = \mathrm{Sigmoid}\big(\Psi_{c_1}(\mathbf{v}_{bs}^{(L)})\big) \in \mathbb{R}^{M}$
	and $\mathbf{c}_2 = \mathrm{Tanh}\big(\Psi_{c_2}(\mathbf{v}_{bs}^{(L)})\big) \in \mathbb{R}^{2M}$.
	The $m$-th entry of $\mathbf{c}_1$ directly yields the amplitude
	coefficient $\eta_{1,m} = c_{1,m} \in [0,1]$, while $c_{2, m} = c_{2, m,1}/\sqrt{2} + \, \jmath \, c_{2,m,2}/\sqrt{2}$.
	\item \textbf{The receive PFV}: Analogously, two linear heads applied to
	the user embedding $\mathbf{v}_k^{(D)}$ produce
	$w_{1,k} = \mathrm{Sigmoid}\big(\Psi_{w_1}(\mathbf{v}_k^{(D)})\big) \in \mathbb{R}$
	and $\mathbf{w}_{2,k} = \mathrm{Tanh}\big(\Psi_{w_2}(\mathbf{v}_k^{(D)})\big) \in \mathbb{R}^2$,
	giving $\eta_{1,k} = w_{1,k}$ and, from $w_{2,k} = w_{2, k,1} + \, \jmath \, w_{2,k,2}$. In
	both cases, the resulting continuous amplitude and phase values are
	finally projected onto the nearest feasible levels of the quantization
	sets $\mathcal{A}$ and $\mathcal{S}$, with $B_\eta$ and
	$B_\phi$ quantization bits, respectively, so that constraint~\eqref{P1f} is
	satisfied.   
\end{itemize}
\vspace{-01.1em}
\subsection{Loss Function Designs}
The network is trained end-to-end, in an unsupervised manner, to
directly maximize the weighted-sum utility of problem $\mathcal{P}_2$
in~(14a) while penalizing violations of the reliability
constraint~(11d), which cannot otherwise be guaranteed by construction
(unlike the power and latency constraints, which are structurally
enforced by the normalization and sigmoid mappings in the readout
layer). Given the achievable rate $R_k$ and DEP $\epsilon_k$ of user $k$
computed from $(\mathbf{p}_k, \{\mathbf{c}_m\}, \mathbf{w}_k, L_k)$
through~(7)--(10), the loss function is defined as $	\mathcal{L} =$
\begin{equation}
 -\Big(\omega\, U_1 + (1-\omega)\, U_2\Big) + \textstyle\sum_{k \in \mathcal{K}} \mu_k\, \mathrm{ReLU}\big(\epsilon_k - \epsilon_{\max}\big),
	\label{eq:loss}
\end{equation}
where $U_1$ and $U_2$ are the normalized sum-rate and reliability
utilities defined in~(12)--(13), and $\mu_k > 0$ is a positive penalty
coefficient that penalizes any user whose DEP exceeds the reliability
target $\epsilon_{\max}$. 

\begin{figure}[t]
	\centering
	\includegraphics[width=0.60\columnwidth]{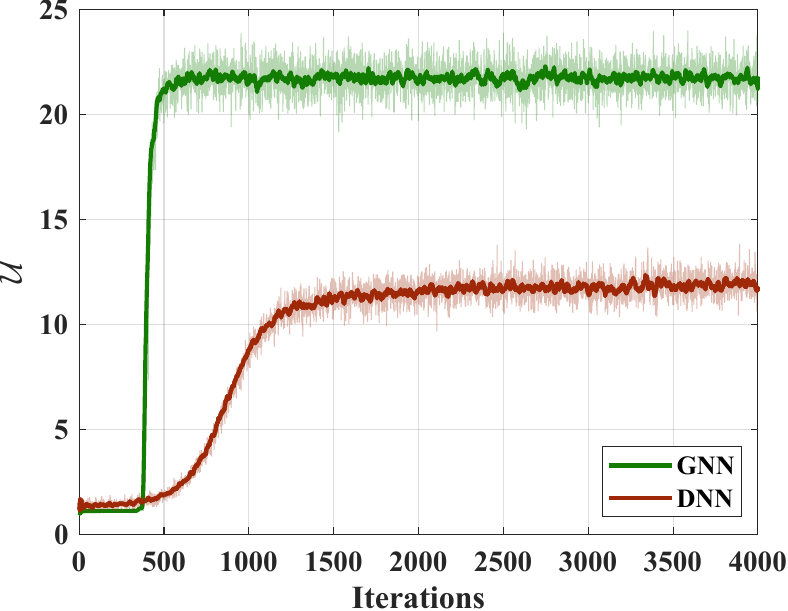}
	\caption{Convergence curve.}
	\label{fig:conv}
\end{figure}
\captionsetup[subfigure]{labelformat=simple, labelsep=colon}
\begin{figure*}[!ht]
	\centering
	\begin{subfigure}[b]{0.24\textwidth} 
	\includegraphics[
	width=\linewidth,
	trim={0 1 1 1},
	clip
	]{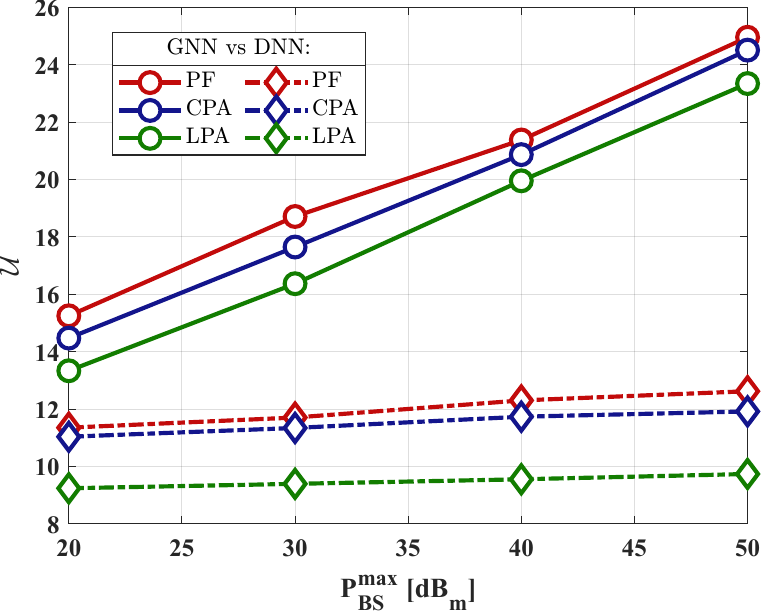}
		\caption{Utility vs. $P_{\mathrm{BS}}^{\max}$}
		\label{fig:power}
	\end{subfigure}\hspace{0.000000001\textwidth} 
	\begin{subfigure}[b]{0.24\textwidth}
		\includegraphics[
		width=\linewidth,
		trim={0 1 1 1},
		clip
		]{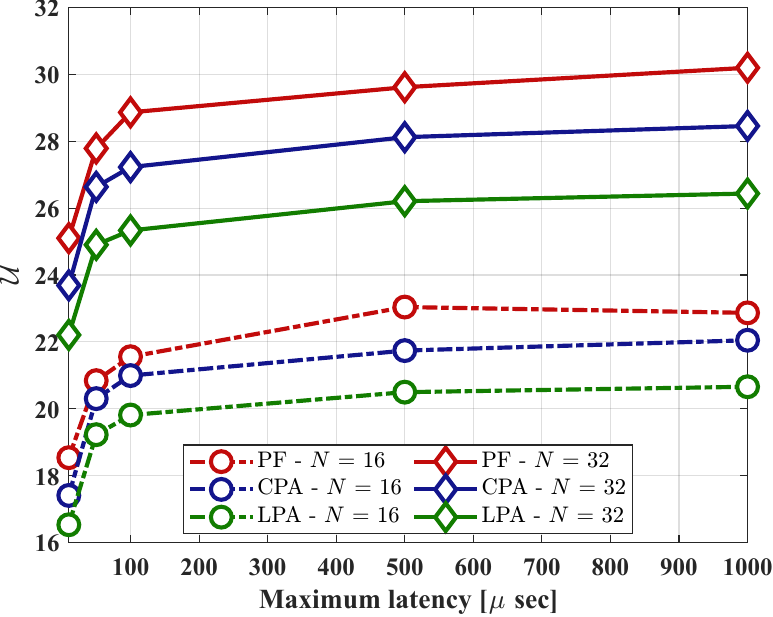}
		\caption{Utility vs. $T_{\max}$.}
		\label{fig:latency}
	\end{subfigure}\hspace{0.000000000001\textwidth}
	\begin{subfigure}[b]{0.24\textwidth}
	\includegraphics[
	width=\linewidth,
	trim={0 1 1 1},
	clip
	]{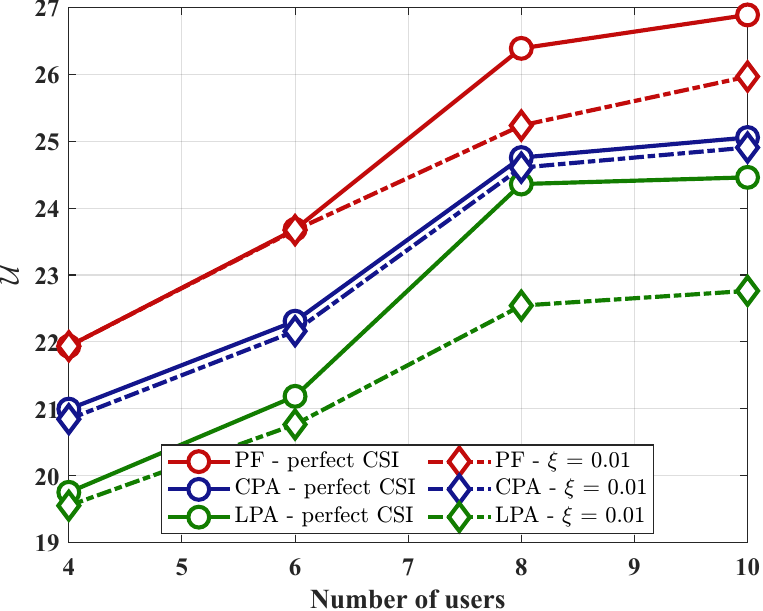}
		\caption{Utility vs. $K$.}
		\label{fig:users}
	\end{subfigure}\hspace{0.000000000001\textwidth}
	\begin{subfigure}[b]{0.24\textwidth}
	\includegraphics[
	width=\linewidth,
	trim={0 1 1 1},
	clip
	]{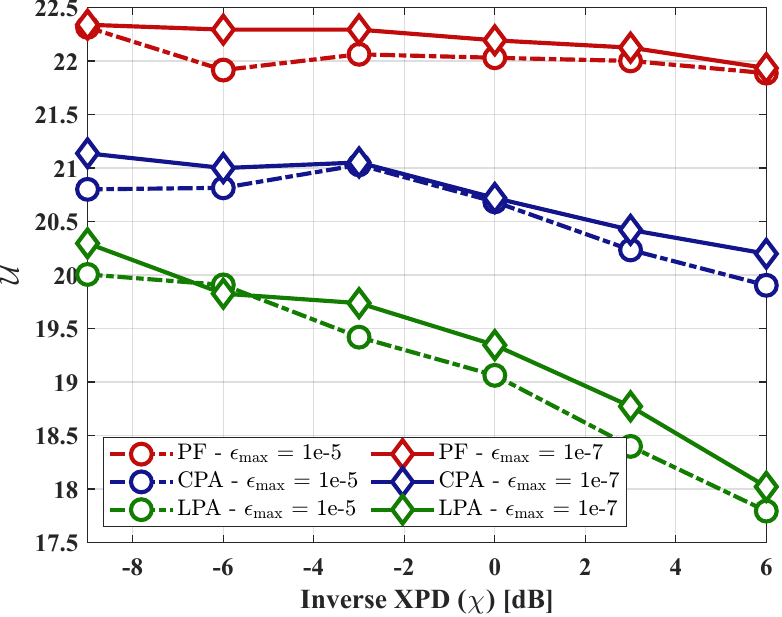}
		\caption{Utility vs. $\chi_{\mathrm{dB}}$.}
		\label{fig:xpd}
	\end{subfigure}	
	\caption{Simulation results.}
	\label{Fig0}
	\vspace{-1.3em}
\end{figure*}
\section{Simulation Results}

In this section, we evaluate the performance of the proposed GNN-based PF framework and compare it with conventional circularly polarized antenna (CPA) and linearly polarized antenna (LPA) configurations. Unless otherwise stated, the key system and simulation parameters listed in Table~\ref{Tab:SimulationParameters} are adopted. The GNN is trained offline and subsequently used for online inference to generate the digital beamforming and polarization-control variables for each channel realization.

\begin{table}[t]
	\centering
	\caption{Key Simulation Parameters}
	\label{Tab:SimulationParameters}
	\setlength{\tabcolsep}{4.5pt}
	\renewcommand{\arraystretch}{1.05}
	\begin{tabular}{ll@{\qquad\quad\quad}ll}
		\toprule
		\textbf{Parameter} & \textbf{Value} &
		\textbf{Parameter} & \textbf{Value} \\
		\midrule
		$f_c$ & $24~\mathrm{GHz}$ &
		$T_{\max}$ & $100~\mu\mathrm{s}$ \\
		
		$C_0$ & $\left(\frac{\lambda}{4\pi}\right)^2$ &
		$D_k$ & $128~\mathrm{bits}$ \\
		
		$\alpha$ & $2.3$ &
		$\epsilon_{\max}$ & $10^{-5}$ \\
		
		$M$ & $16$ &
		$\xi$ & $0.01$ \\
		
		$K$ & $4$ &
		$\chi_{\mathrm{dB}}$ & $-6~\mathrm{dB}$ \\
		
		$P_{\mathrm{BS}}^{\max}$ & $40~\mathrm{dBm}$ &
		$B_\eta$ & $3$ \\
		
		$\sigma^2$ & $-90~\mathrm{dBm}$ &
		$B_\phi$ & $3$ \\
		
		Bandwidth, $W$ & $10~\mathrm{MHz}$ &
		Learning rate & $0.001$ \\
		\bottomrule
	\end{tabular}
	\vspace{-1.2em}
\end{table}

Fig.~\ref{fig:conv} compares the convergence of the proposed GNN and DNN in terms of normalized utility. The GNN converges faster and achieves a higher utility, benefiting from its structure-aware representation of the multi-user optimization problem.

Fig.~\ref{fig:power} illustrates the utility as a function of the maximum BS transmit power $P_{\mathrm{BS}}^{\max}$. The proposed GNN-based PF scheme consistently achieves higher utility than the CPA and LPA baselines over the considered power range. In addition, the GNN-based implementation outperforms the corresponding DNN-based implementation, indicating that explicitly modeling the interactions among users through graph-based message passing can provide additional benefits for the considered multi-user optimization problem.

Fig.~\ref{fig:latency} shows the effect of the maximum allowable latency $T_{\max}$ on the weighted-sum utility for $M=16$ and $M=32$. As $T_{\max}$ increases, the available blocklength $L_k=WT_k$ also increases, which reduces the finite-blocklength penalty and consequently improves the utility. The performance improvement gradually diminishes at larger latency values, resulting in a saturation trend. Increasing the number of BS antennas further improves the utility by providing additional spatial degrees of freedom. Across both antenna configurations, the proposed PF framework maintains a clear performance advantage over the fixed-polarization baselines.

Fig.~\ref{fig:users} evaluates the impact of the number of users under perfect CSI ($\xi=0$) and imperfect CSI ($\xi=0.01$). As expected, imperfect CSI degrades the utility because the estimated channel no longer perfectly represents the actual propagation conditions. Nevertheless, the proposed PF framework maintains a performance advantage over the fixed-polarization baselines as the number of users increases, indicating that the polarization-control mechanism remains beneficial in the presence of channel estimation errors.

Fig.~\ref{fig:xpd} investigates the effect of channel depolarization by plotting the utility against the inverse XPD level for two reliability targets, $\epsilon_{\max}=10^{-5}$ and $10^{-7}$. As the inverse XPD increases, the polarization components become more strongly coupled, leading to increased depolarization and polarization mismatch. The proposed PF framework exhibits a comparatively stable utility performance over the considered inverse-XPD range, whereas the CPA and LPA baselines experience more pronounced performance degradation. This behavior demonstrates the ability of adaptive polarization control to mitigate the impact of channel depolarization.

Overall, the numerical results demonstrate that the proposed GNN-based polarforming framework provides substantial utility gains over fixed-polarization transmission and retains its performance advantage under imperfect CSI and varying channel depolarization levels. The results also highlight the benefit of exploiting the graph structure of the multi-user system for learning the coupled beamforming and polarization-control policy.

\section{Conclusions}

In this paper, we investigated polarization-aware transmission for MU-MISO short-packet URLLC systems under imperfect CSI. Digital beamforming, transmit/receive polarforming, and blocklength allocation were jointly optimized to balance the FBL sum rate and maximum DEP subject to transmit-power, latency, reliability, and discrete polarization-control constraints. The resulting multi-objective problem was scalarized through a normalized weighted-sum utility.

To enable low-complexity online decision-making, we developed a heterogeneous GNN that represents the BS and users as different node types and exploits the interactions among users. The GNN learns a direct mapping from estimated polarized CSI to the digital beamforming and polarization-control variables, thereby avoiding the need to repeatedly solve the underlying optimization problem for each channel realization.

Numerical results showed that the proposed GNN-based polarforming framework improves the weighted-sum utility over conventional CPA and LPA configurations. The proposed approach also maintains a performance advantage under imperfect CSI and exhibits improved robustness to channel depolarization. These results demonstrate the potential of combining polarization-domain adaptation with graph-based learning for short-packet URLLC transmission in the presence of practical CSI imperfections.

\bibliographystyle{IEEEtran}
{\footnotesize
	\bibliography{Polarforming-URLLC}} 

@techreport{3GPP2016,
author      = {{3GPP}},
title       = {Study on Scenarios and Requirements for Next Generation Access Technologies},
institution = {3rd Generation Partnership Project},
number      = {TR 38.913},
version     = {14.3.0},
year        = {2017}
}

@article{Vu2025SPC,
author  = {Thai-Hoc Vu and others},
title   = {Short-Packet Communications: Recent Advances and Research Challenges},
journal = {IEEE Network},
year    = {2025},
note    = {Early Access},
doi     = {10.1109/MNET.2025.3604248}
}

@article{Polyanskiy2010,
author  = {Yury Polyanskiy and others},
title   = {Channel Coding Rate in the Finite Blocklength Regime},
journal = {IEEE Trans. Inf. Theory},
volume  = {56},
number  = {5},
pages   = {2307--2359},
year    = {2010},
doi     = {10.1109/TIT.2010.2043769}
}

@article{Ostman2021URLLC,
author  = {Johan {\"O}stman and Alejandro Lancho Serrano and Giuseppe Durisi and Luca Sanguinetti},
title   = {{URLLC} With Massive {MIMO}: Analysis and Design at Finite Blocklength},
journal = {IEEE Trans. Wireless Commun.},
volume  = {20},
number  = {10},
pages   = {6387--6401},
year    = {2021},
doi     = {10.1109/TWC.2021.3073741}
}

@article{Yeganeh2025LEODRL,
author  = {Rahman Saadat Yeganeh and Hamid Behroozi},
title   = {Energy Efficient {RSMA}-Based {LEO} Satellite Communications Assisted by {UAV}-Mounted {BD}-Active {RIS}: A {DRL} Approach},
journal = {arXiv preprint arXiv:2505.04148},
year    = {2025},
doi     = {10.48550/arXiv.2505.04148}
}

@article{Yeganeh2025ASRIS,
author  = {Rahman Saadat Yeganeh and others},
title   = {QoS Improvement in Multi User Cellular-Symbiotic Radio Network Assisted by Active-{STAR}-{RIS}},
journal = {IEEE Trans. Cogn. Commun. Netw.},
volume  = {11},
number  = {6},
pages   = {3729--3743},
year    = {2025},
doi     = {10.1109/TCCN.2025.3541060}
}

@article{Wu2026FBL,
author  = {Jingchen Wu and Ruiding Hou and Jiaheng Wang and Yongming Huang and Sen Wang and Liang Xia and Jing Jin},
title   = {Finite Blocklength {MIMO} Precoding With Mixed Power and {QoS} Constraints},
journal = {IEEE Commun. Lett.},
volume  = {30},
number  = {3},
pages   = {1235--1239},
year    = {2026},
doi     = {10.1109/LCOMM.2026.3666785}
}

@article{Peng2025URLLC,
author  = {Hongsen Peng and Meixia Tao},
title   = {Two-Timescale Cross-Layer Design for {URLLC} Over Parallel Fading Channels With Imperfect {CSI}},
journal = {IEEE Open J. Commun. Soc.},
volume  = {6},
pages   = {4126--4139},
year    = {2025},
doi     = {10.1109/OJCOMS.2025.3564296}
}

@article{Zhou2024Polarforming,
author  = {Zijian Zhou and others},
title   = {Polarforming for Wireless Communications: Modeling and Performance Analysis},
journal = {IEEE Trans. Wireless Commun.},
year    = {2025},
note    = {Early Access},
doi     = {10.1109/TWC.2025.3524249}
}

@article{Ding2026Opportunities,
author  = {Jingze Ding and others},
title   = {Polarforming for Wireless Networks: Opportunities and Challenges},
journal = {IEEE Commun. Mag.},
volume  = {64},
number  = {3},
pages   = {118--124},
year    = {2026},
month   = {Mar.},
doi     = {10.1109/MCOM.001.2500341}
}

@inproceedings{Zhou2025PS,
author    = {Zijian Zhou and Jingze Ding and Rui Zhang},
title     = {Polarforming Design with Phase Shifter Based Polarization Reconfigurable Antennas},
booktitle = {Proc. IEEE 102nd Veh. Technol. Conf. (VTC2025-Fall)},
year      = {2025},
month     = {Oct.},
doi       = {10.1109/VTC2025-Fall65116.2025.11310254}
}

@ARTICLE{11547205,
	author={Mehrzad, Saba and Ebrahimi, Kamran and Attaran, Ali and Fang, Fang},
	journal={IEEE Wireless Commun. Lett.}, 
	title={Polarforming for {IRS}-Assisted Flexible Intelligent Metasurfaces-Enabled Wireless Communications}, 
	year={2026},
	volume={15},
	number={},
	pages={3576-3580},
	doi={10.1109/LWC.2026.3699597}}

@article{Zhou2025ThreeD,
author  = {Zijian Zhou and Jingze Ding and Jie Xu and Rui Zhang},
title   = {Modeling and Optimization of Three-Dimensional Polarforming for Multiuser Communications},
journal = {J. Commun. Inf. Netw.},
volume  = {10},
number  = {4},
pages   = {364--375},
year    = {2025},
month   = {Dec.},
doi     = {10.23919/JCIN.2025.11357497}
}

@article{Shao2026PAISAC,
author  = {Xiaodan Shao and others},
title   = {Polarforming Antenna Enhanced Sensing and Communication: Modeling and Optimization},
journal = {IEEE J. Sel. Areas Commun.},
volume  = {44},
pages   = {416--431},
year    = {2026},
doi     = {10.1109/JSAC.2025.3610487}
}

@article{Li2026LEOPolarforming,
author  = {Haoyue Li and Li You and Mengyu Qian and Yuanshuo Wang and Huibin Zhou and Xiqi Gao},
title   = {Polarforming and Precoding Design for Massive {MIMO} {LEO} Satellite Communications},
journal = {IEEE Wireless Commun. Lett},
volume  = {15},
pages   = {4045--4049},
year    = {2026},
doi     = {10.1109/LWC.2026.3702745}
}

@article{Li2024GNNMUMISO,
author  = {Yuhang Li and Yang Lu and Bo Ai and Octavia A. Dobre and Zhiguo Ding and Dusit Niyato},
title   = {{GNN}-Based Beamforming for Sum-Rate Maximization in {MU-MISO} Networks},
journal = {IEEE Trans. Wireless Commun.},
volume  = {23},
number  = {8},
pages   = {9251--9264},
year    = {2024},
doi     = {10.1109/TWC.2024.3361174}
}

@article{Le2025GNNSTAR,
author  = {Ha An Le and others},
title   = {Graph Neural Network-Based Active and Passive Beamforming for Distributed {STAR-RIS}-Assisted Multi-User {MISO} Systems},
journal = {IEEE Trans. Commun.},
volume  = {73},
number  = {10},
pages   = {9299--9312},
year    = {2025},
doi     = {10.1109/TCOMM.2025.3558504}
}

@article{Tang2025TwoPhaseGNN,
author  = {Huijun Tang and Jieling Zhang and Zhidong Zhao and Huaming Wu and Hongjian Sun and Pengfei Jiao},
title   = {Joint Optimization Based on Two-Phase {GNN} in {RIS}- and {DF}-Assisted {MISO} Systems With Fine-Grained Rate Demands},
journal = {IEEE Trans. Wireless Commun.},
volume  = {24},
number  = {12},
pages   = {9989--10002},
year    = {2025},
doi     = {10.1109/TWC.2025.3576298}
}
	\end{document}